\documentclass[preprint]{JHEP3}

\def\bea{\begin{eqnarray}}
\def\eea{\end{eqnarray}}
\def\bec{\begin{center}}
\def\ec{\end{center}}

\def\beq{\begin{equation}}
\def\eeq{\end{equation}}

\def\p{\partial}
\def\f{\frac}

\def\f#1#2{\frac{#1}{#2}}
\def\p{\partial}

\preprint{KAIST-TH 02/05 \,\, \,KIAS-P02002} 
\title{Gauge coupling renormalization 
in orbifold field theories}
\author{Kiwoon Choi\\Department of Physics,
Korea Advanced Institute of Science and Technology
Daejeon 305-701, Korea  
\\ E-mail: \email{kchoi@hep.kaist.ac.kr}}
\author{Hyung Do Kim\\School of Physics,Korea Institute for Advanced Study
Seoul 235-010, Korea 
\\ E-mail: \email{hdkim@kias.re.kr}}
\author{Ian-Woo Kim\\Department of Physics,
Korea Advanced Institute of Science and Technology
Daejeon 305-701, Korea  
\\ E-mail: \email{iwkim@hep.kaist.ac.kr}}
\abstract{
We investigate the gauge coupling renormalization 
in orbifold field theory preserving 4-dimensional
$N=1$ supersymmetry in the framework of
4-dimensional effective supergravity.
As a concrete example, we consider the 5-dimensional Super-Yang-Mills theory 
on a slice of ${\rm AdS}_5$. 
In our approach, one-loop gauge couplings
can be determined by the loop-induced axion couplings
and the tree level properties of 4-dimensional effective supergravity
which are much easier to be computed.
}
\begin{document}
%%%%%%%%%%%%%%%%%%%%%%%%%%%%%%%%%%%%%%%%%%%%%%%%%%%%%%%%
\section{Introduction}

Recently higher-dimensional field theories compactified on orbifold
have been proposed as models providing an efficient mechanism
for  symmetry breaking, e.g. the supersymmetry (SUSY) breaking
\cite{Scherk:1978ta,Pomarol:1998sd} 
and/or the grand unified gauge symmetry breaking
\cite{Kawamura:1999nj}.
One can construct realistic grand unified models more
efficiently in such framework \cite{Hall:2001xb,Altarelli}.
Higher-dimensional orbifold models may also lead to a geometric understanding 
of various hierarchical mass scales in particle physics
\cite{Antoniadis:1990ew,Randall:1999ee},
the suppression of some  Yukawa couplings
\cite{Arkani-Hamed:1999dc}, and the 
$b$-$t$ mass ratio \cite{Kim:2001at}.
%%%%% Proper references for Yukawa suppression should be put!!! %%%%%

In this paper, we wish to discuss 
the gauge coupling renormalization in orbifold field theories
preserving the 4-dimensional (4D) $N=1$ supersymmetry
in the framework of 4D effective supergravity (SUGRA).
We are particularly interested in theories with large 
scale hierarchies, for instance
a model in which the Kaluza-Klein (KK) threshold scale 
is significantly lower than the cutoff scale of the theory
so that the KK towers are important for the gauge coupling
renormalization.
In such cases, it is quite convenient to consider
the gauge coupling renormalization in the framework
of 4D effective SUGRA since the 1-loop gauge couplings
can be determined by the {\it loop-induced
axion couplings} and the {\it tree level properties} of 4D effective SUGRA
which are much easier to be computed.
As a concrete example, we will consider 5D SUGRA-coupled 
super-Yang-Mills (SYM) theory on a slice of 5D Anti-de Sitter space (${\rm AdS}_5$) 
\cite{Altendorfer:2000rr,Gherghetta:2000qt,Marti:2001iw,pomarol,Randall:2001gb} 
with four well-seperated  mass scales: the KK scale, the
orbifold length, the AdS curvature, and finally the cutoff scale. 
However, much of the discussions here can be easily extended to
generic higher dimensional orbifold field theories. 
Also the results for a flat supersymmetric 5D geometry
can be obtained from our AdS results by taking the limit that
the AdS curvature becomes zero.

The organization of this paper is as follows.
In section II, we discuss some features
of the supersymmetric gauge theory on a slice of ${\rm AdS}_5$,
including the gauged $U(1)_R$ symmetry and 
also the possible form of gauge coupling renormalization.
In section III, we derive the 4D effective SUGRA of the
5D theory on ${\rm AdS}_5$ and match the renormalized low energy gauge couplings in
5D theory with the moduli-dependent gauge couplings in 4D effective SUGRA.
With this matching, one can determine the 1-loop gauge couplings
by the 1-loop induced axion couplings
and the tree level properties of 4D effective SUGRA.
Section IV is the conclusion.

\section{Supersymmetric gauge theory  on a slice of ${\rm AdS}_5$}

In this section, we discuss some features of the
SUGRA coupled SYM theory on a slice of ${\rm AdS}_5$ which is orbifolded
w.r.t $y\rightarrow -y$ and $y\rightarrow y+2\pi$ where
$y$ is the coordinate of the 5-th dimension.
The model under consideration contains 5D vector multiplets
for YM fields, 5D hypermultiplets for charged matter fields, as
well as the  5D SUGRA multiplet.
To be general, we also assume that there are charged brane matter fields
living on 3-branes  at the orbifold fixed points
$y=0$ and $\pi$.

The action of the SUGRA multiplet is given by
\cite{Altendorfer:2000rr,Gherghetta:2000qt}
%\cite{Cremmer:1980gs}
\bea
\label{5daction1}
S_{\rm sugra} &=& -\f{1}{2}\int d^4x dy \sqrt{-G}\,
M_*^3 \left\{\,
{\cal R}+\bar{\Psi}^i_M\gamma^{MNP}D_N\Psi_{i\,P}+\frac{3}{2}C^{MN}C_{MN}
\right.
\nonumber \\
&&\left.-\frac{1}{4\sqrt{-G}}\epsilon^{MNPQR}B_MC_{NP}C_{QR}
-\f{3}{2}k\epsilon (y)\bar{\Psi}^i_M\gamma^{MN}(\sigma_3)_{ij}
\Psi^j_N\right. 
\nonumber \\
&&\left.-12k^2+\f{\left(\,
\delta(y)-\delta(y-\pi)\,\right)}{\sqrt{G_{55}}}12k+...\,\right\}\,,
\eea
where ${\cal R}$ is the 5D Ricci scalar
for the metric $G_{MN}$,
$C_{MN}=\partial_MB_N-\partial_NB_M$ is the field
strength of the graviphoton $B_M$, $\Psi^i_M$ ($i=1,2$)
are the symplectic Majorana gravitinos, 
$M_*$ is the 5-dimensional
Planck scale, $k$ is the AdS curvature, and the ellipsis
stands for the higher dimensional terms which are not relevant for the present
discussion. 
Here the gravitino kink mass and
the brane cosmological constants are determined by supersymmetry, and
the indices $i,j$ label the fundamental representation
of the $SU(2)_R$ automorphism group of 5D SUGRA, which is raised or lowered
by $\varepsilon^{ij} = \varepsilon_{ij} = (i\sigma_2)^{ij}$ in the NW-SE 
convention. 
The SUGRA multiplet has the standard $Z_2$ boundary condition:
\bea
\label{bc1}
e^A_M(-y)&=&Z^A_BZ^N_Me^B_N(y)\,,
\nonumber \\
\Psi^i_M(-y)&=&Z^N_M(\sigma_3)^i_{~j}\gamma_5\Psi^j_N(y)\,,
\nonumber \\
B_M(-y)&=&-Z^N_MB_N(y)\,,
\eea
where $e^A_M$ is the 5-bein and
$$
Z_M^N=Z_A^B={\rm diag}\,(1,1,1,1,-1).
$$
The above orbifolding breaks the $N=2$ SUSY down to 
$N=1$.
One may break the residual $N=1$ SUSY by imposing a nontrivial
boundary condition of $\Psi^i_M$ for $y\rightarrow y+2\pi$
(the Sherk-Schwarz SUSY breaking). In this paper, we consider
only the orbifolding preserving $N=1$ SUSY, so 
all gravity multiplets are assumed to be periodic under $y\rightarrow y+2\pi$.
The above action for SUGRA multiplet leads to the following
${\rm AdS}_5$ metric as a solution of the equations of motion:
\beq
\label{5dmetric}
ds^2\,=\, e^{-2k|y|R}g_{\mu\nu}dx^\mu dx^\nu+R^2dy^2\,,
\eeq
where $g_{\mu\nu}$ denotes the massless 4D graviton and
$R$ is the orbifold radius.

The action of vector multiplets is given by \cite{Gherghetta:2000qt}
\bea
\label{5daction2}
S_{\rm vector}&=& -\f{1}{2}\int d^4xdy\sqrt{-G}
\left\{\,\f{1}{2g_{5a}^2}F^{aMN}F^a_{MN}+D_M\phi^aD^M\phi^a
+\bar{\lambda}^{ia}\gamma^MD_M\lambda^{a}_i\right.
\nonumber \\
&&+\frac{1}{4\sqrt{-G}g_{5a}^2}
\epsilon^{MNPQR}B_MF^{a}_{NP}F^a_{QR}
+\f{1}{2}k\epsilon(y)\bar{\lambda}^{ia}(\sigma_3)_{ij}\lambda^{ja}
-4k^2\phi^a\phi^a
\nonumber \\
&&+\frac{\delta(y)}{\sqrt{G_{55}}}
\left(\frac{1}{2g_{0a}^2}
F^{a\mu\nu}F^a_{\mu\nu}+4k\phi^a\phi^a+... \right)
\nonumber \\
&&\left.+\f{\delta(y-\pi)}{\sqrt{G_{55}}}\left(
\f{1}{2g_{\pi a}^2}F^{a\mu\nu}F^a_{\mu\nu}-4k\phi^a\phi^a+...\right)\,
+...\,\right\}\,,
\eea
where $F^a_{MN}$ are the YM field strengths, $\lambda^{ia}$ are
the symplectic Majorana gauginos, $\phi^a$ are real scalar fields
spanning a very special target manifold, and the ellipses denote
the higher dimensional terms.
Again the gaugino kink mass and the scalar (bulk and brane) masses are 
determined by supersymmetry.
Here we are interested in the vector multiplets containing
zero mode gauge fields, so the $Z_2$ boundary conditions
are given by
\bea
\label{bc2}
A^a_M(-y)=Z^N_MA^a_N(y),
\quad
\lambda^{ia}(-y)=(\sigma_3)^i_{~j}\gamma_5 \lambda^{ja}(y),
\quad
\phi^a(-y)=-\phi^a(y),
\eea
where all fields in the vector multiplet are assumed to be 
periodic under $y\rightarrow y+2\pi$.

The action of hypermultiplets has the form \cite{Gherghetta:2000qt}
\bea
\label{5daction3}
S_{\rm hyper} &=&
\int d^4x dy \sqrt{-G}\, \left[\,
|D_Mh_I^{i}|^2+i\bar{\Psi}_I\gamma^MD_M\Psi_I+
ic_Ik\epsilon(y)\bar{\Psi}_I\Psi_I \right.
 \nonumber \\
&&\left.+\sum_I \sum_i \left(\,
(c_I^2+\epsilon_i c_I-\f{15}{4}) k^2+
k(3-2\epsilon_i c_I)\f{(\delta(y)-\delta(y-\pi))}{\sqrt{G_{55}}}\right)|h^i_I|^2
+...\,\right]
\eea
where $h_I^i$  are two complex scalar fields
in the $I$-th hypermultiplet spanning a quaternionic target manifold
with the tangent space group $SU(2)_R\times Sp(2n_H)$ ($I=1,.., n_H$), 
$\Psi_I$ are the Dirac fermions 
with kink mass $c_Ik\epsilon(y)$, and $\epsilon_{1,2}=\pm 1$.
Under the
$[U(1)]^{n_H}$ subgroup of
$[SU(2)]^{n_H}\subset Sp(2n_H)$,
the hypermultiplets transform as\footnote{Here we choose
the scalar field fluctuations not being the coordinates of the
quaternionic target manifold, but being a fundamental representation
of the tanget space group $SU(2)_R\times Sp(2n_H)$, which is always possible
with the $4n_H$-beins on the quaternionic  manifold.}
$$
h^i_I\rightarrow e^{i\alpha_I}h^i_I,\quad
\Psi_I\rightarrow e^{i\alpha_I}\Psi_I,
$$
while the $SU(2)_R$ transformation can be read off from the index $i$.
Here the hypermultiplets are allowed to have nontrivial boundary 
conditions for both $y\rightarrow -y$ and $y\rightarrow y+2\pi$:
\bea
\label{bc3}
h_I^i(-y)&=&\omega_I(\sigma_3)^i_{~j}h_I^j(y),\quad
h_I^i(-y+\pi)=\eta_I(\sigma_3)^i_{~j}h_I^j(y+\pi),
\nonumber \\
\Psi_I(-y)&=& \omega_I\gamma_5\Psi_I(y),
\quad \Psi_I(-y+\pi)=\eta_I\gamma_5\Psi_I(y+\pi),
\eea
where $\omega_I=\pm 1$ and $\eta_I=\pm 1$.
Note that $y+\pi\rightarrow -y+\pi$ corresponds to the successive
transformation of $y\rightarrow -y$ and $y\rightarrow y+2\pi$.

To obtain supersymmetric AdS background, we need to gauge $U(1)_R$
symmetry by graviphoton. 
%Then the AdS curvature is related with the 
%$U(1)_R$ gauge coupling $g_R$. 
It has been noted in \cite{Bergshoeff:2000zn} that
the gauge coupling of $B_M$ is required to 
be $Z_2$-odd when the 5D SUGRA model 
contains matter fields and nontrivial brane actions.
%\bea
%g_B=\hat{g}_B\epsilon (y)\,,
%\eea
%where $\hat{g}_B$ is a constant.
Then the $U(1)_R$ gauge transformation and the corresponding
covariant derivative are given by
\bea
&&\Phi\, \rightarrow\, e^{-i\epsilon (y)\Omega(x,y)g_BT_B}\Phi\,,
\nonumber \\
&&B_M\,\rightarrow\, B_M+\p_M\Omega,
\nonumber \\
&& D_M\Phi=\p_M\Phi+i\epsilon (y) g_BT_B B_M\Phi ,
\eea
where $g_B$ is a coupling constant,
$T_B$ is the gauged $U(1)_R$ generator,
and $\Omega$ is a continuous gauge function 
obeying the orbifolding boundary condition
$$
\Omega (y)=-\Omega (-y) =\Omega (y+ 2\pi).
$$
Note that $\Omega(0)=\Omega (\pi)=0$ guarantees that (i)
$D_M\Phi$ has the same gauged $U(1)_R$ transformation as $\Phi$,
(ii) the Chern-Simons terms in the action are invariant
under the gauged $U(1)_R$, (iii) there is no $U(1)_R$-anomaly.
It also means that bulk and brane matter fields at $y=0,\pi$ are all
invariant under the gauged $U(1)_R$.

In order for the $U(1)_R$ covariant derivative to be consistent with the SUSY
transformation, $T_B$ is required to commute with
the $Z_2$ transformation associated with $y\rightarrow -y$.
Without loss of generality, such $U(1)_R$ generator
can be written as\footnote{Note that $T_B$ can always be matched 
with an isometry generator in the quaternionic manifold of hypermultiplet
scalar fields.}  
\bea
g_BT_B=g_RT_{3R}+ g_IT_{3I}\,,
\eea
where $T_{3R}$ is the $U(1)$-generator of $SU(2)_R$
and $T_{3I}$ is the $U(1)$-generator of the $I$-th $SU(2)$
subgroup of $Sp(2n_H)$.
It is then straightforward to find
(see Appendix)
\bea
g_R=-3k\,,\quad g_I=c_Ik\,,
\eea
yielding the following form of covariant derivatives:
\bea
\label{u1rcovariant}
D_Mh^i_I&=&\nabla_Mh^i_I-i\left(\f{3}{2}(\sigma_3)^i_{~j}-c_I\delta^i_{~j}\right)k
\epsilon (y) B_Mh^j_I\,,
\nonumber \\
D_M\Psi_I&=& \nabla_M\Psi_I+ic_Ik\epsilon (y) B_M\Psi_I\,,
\nonumber \\
D_M\lambda^{ia}&=&\nabla_M\lambda^{ia}-i\f{3}{2}(\sigma_3)^i_{~j}
k\epsilon(y)B_M\lambda^{aj}\,,
\eea
where $\nabla_M$ denotes the covariant derivative containing other gauge fields
including the spin connection.

In addition to  $S_{\rm sugra}$, $S_{\rm vector}$
and $S_{\rm hyper}$, there can be additional brane actions 
involving the brane fields as well as the bulk fields at the
fixed points.
Such brane actions are required to be invariant
under the unbroken $N=1$ SUSY generated by the 
Killing spinor on ${\rm AdS}_5$
which will be discussed in Appendix.
Also the gauged $U(1)_R$ enforces that
$B_M$ can appear in the brane actions only
through $C_{\mu 5}=\p_\mu B_5-\p_5 B_\mu$, not through the covariant
derivative of matter field. 
As a result, the brane actions do not have
any non-derivative coupling of $B_5$
in the field basis of $S_{\rm vector}$ and $S_{\rm hyper}$.

Let us now consider the mass scales involved in the model.
Generically, orbifold field theories on a slice of ${\rm AdS}_5$
have the KK scale,
i.e. the scale where the massive KK modes start to appear and/or the KK level
spacing, given by
\beq
M_{KK}\simeq \frac{\pi k}{e^{\pi k R}-1}.
\label{kkscale}
\eeq
In the limit of large AdS curvature $\pi kR\gg 1$,
the KK scale is exponentially suppressed, $M_{KK} \simeq \pi k e^{-\pi k R}$,
so we have the mass scale hierarchies\footnote{
Our orbifold field theory has many scalar fields,
e.g. scalar fields in 5D vector or hypermultiplets, which can
have nonzero vacuum expectation values (VEV)
$\langle\sigma\rangle$ in general.
In case that $\langle\sigma\rangle\gg \mu$, 
we should take into account this additional
mass scale in the analysis.
Here we assume that there is no such mass scale,
so no more mass scales other than $M_{KK}$, $1/R$ and $k$ between
$\mu$ and $M_*$.}
\beq
\label{hierarchy}
\mu\, \ll\, M_{KK}\,\ll \,1/R \,\ll\,k\,\ll\, M_*,
\eeq
where $\mu$ denotes the low energy scale for
currently available experiments.
In the other limit with small AdS curvature $\pi kR\ll 1$,
the geometry is (approximately) flat and the KK
scale  is given by
\beq
M_{KK}\simeq 1/R\,,
\eeq
for which the scale hierarchies are given by
\beq
\mu\, \ll\, M_{KK}\,\simeq\, 1/R\, \ll\, M_*.
\eeq

The low energy couplings
of gauge field zero modes in ${\rm AdS}_5$ would appear as a 
dimensionless function of the involved
mass scales $M_{KK}, R, k, M_*$ as well as of the bare
couplings $g_{5a}^2, g_{0a}^2, g_{\pi a}^2$
\cite{pomarol,Randall:2001gb}.
(Here we assume that $M_*$ corresponds to the cutoff scale of the model.)
At tree level, the 4D gauge couplings $g_a^2$ 
are simply given by
$$
\label{tree}
\left(\frac{1}{g_a^2}\right)_{\rm tree}=
\frac{\pi R}{g_{5a}^2}+\frac{1}{g_{0a}^2}+\f{1}{g_{\pi a}^2} \, .
$$
%\footnote{
%Note that in principle the cutoff scale of the model can be
%significantly lower than $M_*$.}
At 1-loop order, there can be two types of
quantum corrections: those which are power-law dependent
on the involved  energy scales and the others which are
logarithmic in scales.
As for the power-law dependent part, it is 
dominated by the contribution from the cutoff scale $M_*$, 
while the logarithmic part receives equally important
contributions from all scales.
When $\pi kR\gg 1$ so that we have the mass hierarchy (\ref{hierarchy}) with
$M_{KK}\approx \pi k e^{-\pi kR}$,
writing the dimensionless 1-loop 4D gauge coupling at
low energy scale $\mu\ll M_{KK}$ in terms of 
the involved mass scales in a manner having sensible limiting behavior 
at large $R$,
we find that $g_a^2(\mu)$ can be generically
written as\footnote{
Here we consider the case with $M_*\gg k$ to see the gauge coupling
renormalization proportional
to $\ln(M_*/k)$. However in most of the
practical applications of the model, one assumes
$k\sim M_*$. The gauge coupling renormalization in such case
can be obtained from our result by simply ignoring $\ln (M_*/k)$
}.
\bea
\label{4dcoupling}
&&\frac{1}{g_a^2(\mu)}\,=\,
\left[\,\frac{1}{g_{5a}^2(M_*)}
+\frac{\gamma_a}{8\pi^3}M_*\,\right]\pi R
+\frac{1}{g_{0a}^2(M_*)}+\f{1}{g_{\pi a}^2(M_*)}+\f{C_a}{8\pi^2}
\nonumber \\
&&\quad\quad
+\frac{b^{\prime\prime\prime}_a}{8\pi^2}\ln\left(
\frac{M_*}{k}\right)+
\frac{b^{\prime\prime}_a}{8\pi^2}\ln\left(kR\right)
+\frac{b^{\prime}_a}{8\pi^2}\ln\left(\frac{1}{M_{KK}R}\right)
+\frac{b_a}{8\pi^2}\ln\left(\frac{M_{KK}}{\mu}\right)
\,,
\eea
where $C_a$ are some constants which do {\it not}
depend on any of $\mu, M_{KK}, R$ and $k$ at one-loop approximation,
and  $b_a$ are the conventional
one-loop beta function coefficients receiving
the contribution only from the massless 4D  modes at scales below $\mu$.
The other coefficients $b^{\prime}_a, b^{\prime\prime}_a, b^{\prime\prime\prime}_a$
and $\gamma_a$ receive the contribution
from the KK towers.
Among them,  $b^{\prime}_a,b^{\prime\prime}_a$  and $b^{\prime\prime\prime}_a$
can be unambiguously calculated within the orbifold field theory
as they reflect the infrared property of the model
below the cutoff scale $M_*$, while $\gamma_a$ are uncalculable
as they reflect the unknown UV physics around $M_*$.
We stress that the logarithms  $\ln (M_I/M_J)$ in (\ref{4dcoupling})
for $\{M_I\}=\{M_*,k,1/R,M_{KK}\}$ can {\it not} 
be interpreted solely as the coupling running between $M_I$ and $M_J$. 
There can be some contribution from the coupling running, particularly
from the running localized at the orbifold fixed points. 
However they include also
the finite KK threshold corrections.
So (\ref{4dcoupling}) should be interpreted
as one simple way to reorganize the whole quantum corrections
including both the KK threshold effects and the running effects.

In case with large scale ratios given by (\ref{hierarchy}),
the large logs of 
$b_a,b^{\prime}_a$ and $b^{\prime\prime}_a$
provide important corrections to the 4D gauge couplings.
As for $C_a$ which are independent of mass scales,
they are generically of order unity, so
subleading compared to the large-logs.
One can also make a strong coupling assumption\cite{Chacko:1999hg} 
on the uncalculable bare brane couplings
\beq
\label{strong}
\frac{1}{g_{0a}^2(M_*)}\,\sim \,\f{1}{g_{\pi a}^2(M_*)}
={\cal O}\left(\frac{1}{8\pi^2}\right)\,,
\eeq
and then they are  also subleading compared to the large-logs.

As for the power-law running corrections, their coefficients
$\gamma_a$  highly depend on the way of UV cutoff.
Of course, if the bulk gauge group is a simple group, 
$\gamma_a$ will be $a$-independent.
In other cases that the bulk gauge group is not unified,
one may compute $\gamma_a$ in certain regularization scheme
and argue that different 5D gauge couplings rapidly approach 
to each other (when the scale is increased) due to the power-law 
running governed by $\gamma_a$ \cite{Dienes:1998vh}. 
However, power-law running can not be considered as a calculable property of 
orbifold field theory since it is highly sensitive
to the unknown UV physics \cite{ACG,Contino}.
This can be easily noticed by changing the cutoff
$M_* \rightarrow c_a M_*$ for $a$-dependent constants
$c_a$ which are generically of order unity.
This change of cutoff leads to $\gamma_a\rightarrow c_a\gamma_a$
and represents the effects of unknown threshold
effects at $M_*$.
As usual, the cutoff-scheme dependence of $\gamma_a$ should be
cancelled by the cutoff scheme dependence of 
the corresponding Wilsonian couplings $g^2_{5a}(M_*)$.
It is thus not meaningful to split the 
power-law running part from the bare coupling $1/g_{5a}^2$
in orbifold field theory.
Rather, one has to consider 
the cutoff-scheme independent combination
\beq
\label{kappa}
\kappa_aM_* \equiv \frac{1}{g_{5a}^2(M_*)}+
\frac{\gamma_a}{8\pi^3}M_*.
\eeq

Summarizing the above discussions, when we have the scale hierachy
(\ref{hierarchy}), the 1-loop low energy
gauge couplings  in ${\rm AdS}_5$  can be written as
\bea
\f{1}{g_a^2(\mu)}=\kappa_a \pi M_*R+\f{1}{8\pi^2}\Delta_a+{\cal O}
\left(\f{1}{8\pi^2}\right), 
\eea
where 
\bea
\Delta_a=b_a^{\prime\prime\prime}\ln(M_*/k)+b_a^{\prime\prime}\ln(kR)
+b^{\prime}_a\ln(1/M_{KK}R)+b_a\ln(M_{KK}/\mu)\,
\eea
for $\mu\ll M_{KK}\approx \pi k e^{-\pi kR}$.
Here the 4D momentum $\mu$ is measured in the metric frame of 
massless 4D gravition, and
$\kappa_a$ are uncalculable bare parameters associated
with the bare 5D gauge couplings $g_{5a}^2$.
The other uncalculable
bare brane couplings $g_{0a}^2,g_{\pi a}^2$
are assumed to give subleading corrections
of order $1/8\pi^2$.
The 1-loop correction $\Delta_a$ include both the KK threshold
effects and the running effects which can be unambiguously 
calculated  within the orbifold field theory.
In the next section, we will calculate $\Delta_a$
using the 4D effective SUGRA of 5D theory on ${\rm AdS}_5$.

\section{Matching with 4D effective SUGRA}

In this section, we derive the 4D effective SUGRA
of the 5D SYM theory on a slice of ${\rm AdS}_5$, and
calculate the low energy gauge coupling (\ref{4dcoupling})
in the framework of 4D effective SUGRA.
Generic 4D SUGRA action is determined by the real K\"ahler
potential $K$, the holomorphic
gauge kinetic function $f_a$ and the holomorphic
superpotential $P$.
In superspace, the action  takes the form
\beq
\label{4dactioninsuperspace}
\int \,d^4xd^4\theta  \,\left\{-3\exp \left(-\frac{K}{3}\right)
\right\}
+\left[
\int \,d^4xd^2\theta \, 
\left(\,\frac{1}{4} f_{a}\,W^{a\alpha}W^a_{\alpha}+P\,\right)
+\mbox{h.c}\,\right]
\eeq
where the gravity multiplet fields
are set to their Poincare invariant VEV.
The gauge kinetic function $f_a$ determines 
the 4D Wilsonian gauge coupling ($g_{Wa}$)
and the vacuum angle ($\Theta_{a}$) as
\beq
f_a\,=\,\frac{1}{g_{Wa}^2}+i\frac{\Theta_a}{8\pi^2}\,,
\nonumber
\eeq 
and the K\"ahler potential  can be expanded in powers of 
gauge-charged chiral superfields:
\beq
K\,=\, K_0(T,T^*)+Z_{\Phi}(T,T^*)\Phi^*e^{-V}\Phi+...,
\eeq
where $T$, $\Phi$, and $V$ denote the gauge-singlet moduli superfields, 
gauge-charged chiral matter superfields, and  vector gauge superfields,
respectively.

It has been found in \cite{Novikov:uc}
a relation between the 
beta function and the anomalous dimension in 4D $N=1$ supersymmetric 
gauge theory, which is exact in perturbation theory.
Using this relation, one can express
the low energy one-particle-irreducible (1PI) gauge couplings in terms of 
$f_a$ and the K\"ahler metric
$Z_{\Phi}$ of charged matter fields. 
This procedure can be generalized to find a relation
between the moduli-dependent 1PI gauge coupling and the moduli-dependent
Wilsonian couplings in 4D SUGRA \cite{Kaplunovsky:1994fg}.
For {\it moduli-independent} UV regulator, the super-Weyl and K\"ahler
invariance of 4D SUGRA determines
the 1PI gauge coupling as\footnote{
This relation was confirmed by an explicit computation
in \cite{Kaplunovsky:1994fg} at 1-loop order.} 
\bea
\label{4dsugracoupling}
\frac{1}{g^2_a(\mu)}=
\mbox{Re}(f_a)+\frac{1}{8\pi^2}\Bigg[&&
b_a\ln\left(\frac{M_{\rm Pl}}{e^{-K_0/6}\mu}\right)
-\sum_{\Phi}T_a(\Phi)\ln\left(\frac{{Z}_{\Phi}}{e^{K_0/3}}
\right) \nonumber \\
&&+T_a({\rm Adj})\ln\left(\frac{1}{g_a^2(\mu)}\right)\Bigg],
\eea
where $M_{\rm Pl}$ is the 4D Planck scale of the metric $g_{\mu\nu}$
which is used to measure the external momentum $\mu^2=-g^{\mu\nu}\p_\mu\p_\nu$,
$T_a(\Phi)={\rm Tr}([T_a(\Phi)]^2)$ is the Dynkin index
of the gauge group representation $\Phi$,
and $b_a$ is the one-loop beta function coefficient:
\bea
b_a\,=\, -3T_a({\rm Adj})+\sum_{\Phi}T_a(\Phi)\,.
\nonumber
\eea
Note that the moduli-dependence of $M_{\rm Pl}$ differs in different
metric frame.
For instance, in the 4D superconformal frame in which the action is given by
(\ref{4dactioninsuperspace}), the moduli-dependence of $M_{\rm Pl}$ 
is given by
$e^{-K_0/6}$, while in the 4D Einstein frame which is obtained from
the superconformal frame after the Weyl scaling $g_{\mu\nu}
\rightarrow e^{K_0/3}g_{\mu\nu}$, $M_{\rm Pl}$ is moduli-independent.
When one matches the 4D SUGRA coupling (\ref{4dsugracoupling}) 
with the 1PI coupling computed in underlying orbifold field theory, 
one has to use the same metric 
frame as well as the same {\it moduli-independent} regulator \cite{Choi:2002ps}. 
In the one-loop approximation, 
the above low energy gauge couplings can be written as
\bea
\label{1loopcoupling}
\frac{1}{g_a^2(\mu)}=
{\rm Re}(f_a)+\frac{1}{8\pi^2}\Bigg[&&
b_a\ln
\left(\frac{M_{\rm Pl}}{e^{-K_0/6}\mu}\right)
%\nonumber \\
-\sum_{\Phi}T_a(\Phi)\ln\left(\frac{Z_{\Phi}^{(0)}}{e^{K_0/3}}\right)
\nonumber \\
&&+T_a({\rm Adj})\ln\left({\rm Re}(f_a)\right)
\,\Bigg],
\eea
where $Z^{(0)}_{\Phi}$ is the tree level K\"ahler metric of $\Phi$.

Using (\ref{1loopcoupling}), one can find the 1-loop gauge coupling
in 5D gauge theory on a slice of ${\rm AdS}_5$ by computing 
$f_a, K_0$ and $Z^{(0)}_\Phi$ of the 4D effective SUGRA.
As was noted in the previous section, all scale hierarchies in 5D theory
are determined by the orbifold radius $R$, so the dominant 1-loop renormalization
can be determined by computing the $R$-dependence of the
4D effective SUGRA. 
The only part of (\ref{1loopcoupling}) which
is undetermined by tree-level analysis is the 1-loop 
threshold correction to $f_a$.
However due to the holomorphicity of $f_a$,
in our case of 5D theory on ${\rm AdS}_5$, 
the 1-loop piece of $f_a$ is determined  by the loop-induced 
axion coupling which can be easily computed by
using the chiral anomaly structure of the theory.

In order to derive the 4D effective SUGRA action of the 5D theory on
${\rm AdS}_5$, it is convenient to rewrite the 5D actions 
$S_{\rm vector}$ and $S_{\rm hyper}$ in terms of the $N=1$ superfields of
unbroken SUSY \cite{Nima}.
This procedure involves a $R$ and $B_5$-dependent field redefinition
which would generate a new piece of brane action through the chiral anomaly.
The Killing spinor generating the unbroken $N=1$ global SUSY
is given by (see Appendix)

\bea
e^{-\frac{1}{2} k |y| ( R- 3 iB_5) }\eta, 
\nonumber 
\eea
where $\eta$ is a constant 4D Weyl spinor.
It is then straightforward to see that the 5D vector multiplet
$\{\phi^a, A^a_M, \lambda^{ia}\}$ can be decomposed into
an $N=1$ vector superfield $V^a$ and 
a chiral superfield $\chi^a$ whose component fields are given by
$\{A^a_\mu,\tilde{\lambda}^a\}$ and $\{\tilde{\phi}^a+iA_5, 
\tilde{\zeta}^{a}\}$, respectively, where
\bea
\tilde{\lambda}^a(x,y)&=&e^{-\frac{3}{2}k|y|(R+iB_5)}
\lambda^a(x,y),\nonumber\\
\tilde{\zeta}^{a}(x,y)&=&Re^{-\f{1}{2} k|y|(R-3iB_5)}
\zeta^{a}(x,y),\nonumber \\
\tilde{\phi}^a (x,y) &=& R \phi^a (x,y). \label{vectorredef}
\eea   
for $\lambda^a=\frac{1}{2}(1+\gamma_5)\lambda^{1a}$ and
$\zeta^{a}=\frac{1}{2}(1+\gamma_5)\lambda^{2a}$.
The 5D hypermultiplet $\{h^i_I,\Psi_I\}$ can be similarly
decomposed into two $N=1$ chiral superfields $H_I$ and 
$H^c_I$ whose component fields are given by 
$\{\tilde{h}^1_I,\tilde{\psi}_I\}$ and $\{\tilde{h}^{2*}_I,
\tilde{\psi}^c_I\}$, respectively, where
\bea
\tilde{h}^1_I (x, y ) &=&e^{-\left(\frac{3}{2}-c_I\right) k|y| (R+iB_5)} 
h^1_I (x, y) \nonumber \\
\tilde{h}^2_I (x, y ) &=&e^{\left(\frac{3}{2}+c_I\right) k|y| (-R+iB_5)} 
h^2_I (x, y) \nonumber \\
\tilde{\psi}_I(x,y)&=&e^{-k|y|((2-c_I)R-i c_I B_5 )}\psi_I(x,y),\nonumber\\
\tilde{\psi}^c_I(x,y)&=&e^{ -k|y|((2+c_I)R+ic_I B_5 )}\psi^c_I(x,y),
\label{hyperredef}
\eea
for $\psi_I=\frac{1}{2}(1+\gamma_5)\Psi_I$ and $(\psi^c_I)^*=
\frac{1}{2}(1-\gamma_5)\Psi_I$.

On each brane, there can be a brane action
containing $N=1$ supersymmetric ($B_5$-independent) gaugino-scalar-fermion
interactions $\phi_{\rm br}^*\psi_{\rm br}
\lambda^a$ and/or $\phi_{\rm br}^*\psi_{\rm br}\lambda_{\rm br}$,
where $\{\phi_{\rm br}, \psi_{\rm br}\}$ 
and $\{A^\mu_{\rm br},\lambda_{\rm br}\}$ are the
$N=1$ matter and gauge multiplets, respectively, living
only on the brane.
In order to rewrite such $U(1)_R$-invariant
brane actions in $N=1$ superspace, one needs also to redefine
the brane fermions as follows:
\bea
\label{braneredef}
\tilde{\psi}_{\rm br} &=& e^{-\frac{1}{2} k \left|y_{\rm br}\right| ( R - 3 i 
B_5 )} \psi_{\rm br}, \nonumber\\
\tilde{\lambda}_{\rm br} &=& e^{-\frac{3}{2}  k \left|y_{\rm br}\right|
 (R+i B_5)} \lambda_{\rm br}, 
\eea
where $y_{\rm br}=0$ or $\pi$ 
denotes the location of the brane where the field lives on. 
Then the $N=1$ chiral brane superfield $Q_{\rm br}$
is given by the redefined component fields $\{\phi_{\rm br},
\tilde{\psi}_{\rm br}\}$ and the $N=1$ vector brane superfield
$V_{\rm br}$ by $\{ A^\mu_{\rm br},\tilde{\lambda}_{\rm br}\}$.

After the above field redefinitions, the bulk actions 
$S_{\rm vector}$ and $S_{\rm hyper}$ can be written in $N=1$ superspace
\cite{Marti:2001iw,Nima} as follows
\bea \label{ssaction} S_{\rm bulk}&=&\int d^5x \,\left[\,\int d^4\theta
\,\left\{\frac{1}{2}(T+T^*)e^{-(T+T^*)k|y|}\left(
M_*^3+M_* e^{(\frac{3}{2}-c_I)(T+T^*)k|y|}H_I^*e^{-V}H_I
\right.\right.\right. \nonumber \\
&&\left.\left.+M_* e^{(\frac{3}{2}+c_I)(T+T^*)k|y|}H^c_Ie^VH^{c*}_I)\right)
+2\kappa_a M_* \frac{e^{-(T+T^*)k|y|}}{T+T^*}
(\partial_y V^a-\frac{1}{\sqrt{2}}(\chi^a+\chi^{a*}))^2\right\}
\nonumber \\ && \left. +\int d^2\theta\,\left\{\,
\frac{\kappa_a M_* }{4}T\,W^{a\alpha}W^a_{\alpha}
+H^c(\partial_y-\frac{1}{\sqrt{2}}\chi)H +h.c.\,\right\}
\right]\,,\eea
where $T$ is the radion superfield given by
$$
T=R+iB_5+
\theta \psi_5+\theta^2 F_T\,,
$$
for the fifth-component of the graviphoton ($B_5)$
and $\psi_5=\frac{1}{2}(1+\gamma_5)\Psi_5^{i=2}$.
Here we included the pure radion action
coming from $S_{\rm sugra}$.
The brane actions can be also written in $N=1$  superspace as
\bea
S_{\rm brane}&=&\int_{y=0} d^4x \,\left[\int
d^4\theta\, Q_{UV}^*e^{-V}Q_{UV} +\int d^2\theta \,
\left(\frac{1}{4\hat{g}_{UV}^2}W^{\alpha}_{UV}W_{UV\,\alpha}+
P_{UV}(Q_{UV},H_I)\right)
\right] \nonumber \\ &&+\int_{y=\pi} d^4x\left[\,\int d^4\theta\,
e^{-k\pi (T+T^*)}Q_{IR}^*e^{-V} Q_{IR} 
 +\int d^2\theta
\left(\,\frac{1}{4\hat{g}_{IR}^2}W^{\alpha}_{IR}W_{IR\,\alpha} \right.\right.
\nonumber \\
&&\left.\left.\quad\quad\quad\quad\quad\,+\,e^{-3k\pi
T}P_{IR}(Q_{IR},e^{\left(\frac{3}{2}-c\right) \pi k T} H_I)\,\right)\,\right]
+\hbox{h.c.}\,, \label{braneaction}
\eea
where $Q_{UV}$ and $Q_{IR}$ are the brane chiral superfields
living on the UV ($y=0$) and IR ($y=\pi$) brane, respectively, and
$W_{UV}$ and $W_{IR}$ are the chiral spinor superfields
for the brane vector superfields $V_{UV}$ and $V_{IR}$.

At classical level, the superfield action $S_{\rm bulk}+S_{\rm brane}$
describes the same theory as the component field action before the field
redefinitions (\ref{vectorredef}), (\ref{hyperredef}) and (\ref{braneredef}).
However at quantum level, we should 
include the anomaly terms induced by these field redefinitions at one-loop order.
Since there is no chiral anomaly in 5D bulk, anomalies appear only at the 
orbifold fixed points \cite{Arkani-Hamed:2001is,Scrucca:2001eb}, which 
can be easily calculated to be
\bea
S_{\rm anomaly}=\int
d^5x \int d^2\theta
&&\left[\,\left\{\frac{3}{4}T_a(\lambda^a) \left(\delta(y)+\delta(y-\pi)\right)
-\f{1}{2} c_I T_a(\Psi_I)
\left(\omega_I\delta(y)+\eta_I\delta(y-\pi)\right) \right.\right. \nonumber\\
&& \left.
-\frac{3}{2}T_a(\psi_{IR})
\delta (y-\pi)\,\right\}\frac{k|y|T}{16\pi^2}
W^{a\alpha}W^a_{\alpha}\nonumber \\
&& \left.
+\frac{3}{2}\left(T_{IR}(\lambda_{IR})-T_{IR}(\psi_{IR})\right)
\delta(y-\pi)\frac{k\pi T}{16\pi^2}W^{\alpha}_{IR}W_{IR\,\alpha}
\,\right]
. \label{anomaly}
\eea

It is now straightforward to derive the tree level 4D effective action 
from the above 5D actions in $N=1$ superspace.
The bulk superfields $T$ and $W^a_{\alpha}$ have
($y$-independent) zero modes as they have $(+,+)$ boundary condition 
under $y\rightarrow -y$ and
$y+\pi\rightarrow -y+\pi$.
On the other hand, $H_I$ can have  ($y$-independent) zero mode 
only when $(\omega_I,\eta_I)=(+,+)$, while $H_I^c$ does
only when $(\omega_I,\eta_I)=(-,-)$.
By integrating over $y$, one easily finds the following
radion K\"ahler potential $K_0$, the tree level K\"ahler metric
$Z^{(0)}_{H_I}$, and the tree level gauge kinetic
function $f^{(0)}_a$ for the massless 4D modes of
$T$, $H_I$ or $H^c_I$, and $W^a_{\alpha}$:
\bea
&&e^{-K_0/3}\,=\,(1-e^{-\pi k(T+T^*)})\frac{M_*^3}{k}\,=\, M_{\rm Pl}^2\,,
\nonumber \\
&&e^{-K_0/3}Z^{(0)}_{H_I}\,=\, \frac{M_*}{(\frac{1}{2}- \omega_I c_I)k}
(e^{(\frac{1}{2}-\omega_I c_I)\pi k(T+T^*)}-1)\,,
\nonumber \\
&& f^{(0)}_a=\kappa_a \pi M_*T.
\label{hyperkahler}
\eea
Here we ignored the contribution to $f^{(0)}_a$ from
the fixed point gauge couplings $1/g_{ia}^2$ based on
the strong coupling assumption (\ref{strong}).
The contribution to $Z^{(0)}_{H_I}$ from 
the fixed point kinetic terms of bulk fields are ignored also
by the similar reasoning.
The tree-level K\"ahler metrics and gauge kinetic functions
of brane superfields are also easily
obtained to be
$$
 e^{-K_0/3}Z^{(0)}_{Q_{UV}}\,=\, 1\,,\quad
e^{-K_0/3}Z^{(0)}_{Q_{IR}}\,=\, e^{-\pi k(T+T^*)}\,,
$$
\bea
f^{(0)}_{IR}\,=\,\frac{1}{\hat{g}_{IR}^2}\,,
\quad 
f^{(0)}_{UV}\,=\, \frac{1}{\hat{g}_{UV}^2}\,.
\label{branekahler}
\eea

The $T$-dependent one-loop threshold corrections 
to gauge kinetic functions
can be determined by the loop-induced  
axion ($B_5$) couplings to gauge fields: 
$B_5 \epsilon^{\mu\nu\rho\sigma} F^a_{\mu\nu} F^a_{\rho\sigma}$.
There can be two sources of such axion couplings.  
One is $S_{\rm anomaly}$ and the other is the one-loop threshold
effects of massive KK modes. A nice feature our field basis
is that $B_5$ does not have any non-derivative coupling in $S_{\rm bulk}$ 
other than the Chern-Simons coupling. 
As a result, in our field basis,
integrating out the massive KK modes 
does not generate any $B_5$-coupling to gauge fields.
Then the $T$-dependent one-loop corrections to gauge kinetic functions 
are entirely given by $S_{\rm anomaly}$,
yielding:
\bea
\label{fa1loop}
&&\Delta f_{a}\,=\, -\frac{3}{8\pi^2}\left(
T_a(Q_{IR})-\frac{1}{2}T_a({\rm Adj})+\frac{1}{3}\eta_I c_IT_a(H_I)\right)
k\pi T\,, \nonumber \\ 
&& \Delta f_{IR} \,=\, 
\frac{3}{8\pi^2}\left( T_{IR}({\rm Adj})- T_{IR}(Q_{IR}) \right)
k\pi T\,, \nonumber \\
&& \Delta f_{UV} \,=\, 0. 
\eea

With (\ref{1loopcoupling}), (\ref{hyperkahler}), 
(\ref{branekahler}) and (\ref{fa1loop}), 
the low energy bulk gauge couplings  are given by
\bea
\label{4dbulkcoupling}
\frac{1}{g_a^2(\mu, k,R)} &=& \kappa_a\pi M_* {\rm Re}(T)
+\frac{1}{8\pi^2}\left[
T_a(Q_{UV}) \ln \left(\frac{M_*}{\mu}\right)
+T_a(Q_{IR})\ln \left(\frac{M_* e^{-\f{1}{2}\pi k (T+T^*)}}{\mu}\right) 
\right.\nonumber\\
&& -\,T_a({\rm Adj})\left\{\,3\ln\left( \frac{M_*}{\mu}\right)
-\frac{3}{4}\pi k (T+T^*) -\ln(M_* (T+T^*))\,\right\}\,
\nonumber \\
&&+\sum_{\omega_I = \eta_I} T_a(H_I)\left\{\,\ln\left(\frac{k}{\mu}
\right)
- \ln \left(\frac{e^{\left(\f{1}{2}-\omega_I c_I\right)\pi k(T+T^*)}-1}
{\pi(1-2 \omega_I c_I)}\right)\,\right\} \nonumber \\
&&\left.
-\,\sum_I\,\,\f{1}{2}T_a(H_I)\,\eta_I\,c_I\,\pi k (T+T^*)\,\right]\,,
\eea
where  $\mu^2=-g^{\mu\nu}\partial_\mu\partial_\nu$ is the external
4D momentum below the KK threshold scale.
Here $\sum_I$ denotes the summation over all hypermultiplets, while 
$\sum_{\omega_I =\eta_I}$ denotes the summation over the hypermultiplets
having a zero mode. 
The brane gauge couplings at low energies can be similarly
obtained as
\bea
\label{branecoupling}
\frac{1}{g_{UV}^2(\mu,k,R)}&=&\frac{1}{\hat{g}_{UV}^2}+\frac{b_{UV}}{16\pi^2}
\ln\left(\frac{M_*^2}{\mu^2}\right) \,,
\nonumber \\
\frac{1}{g_{IR}^2(\mu,k,R)}&=& \frac{1}{\hat{g}_{IR}^2}+
\frac{b_{IR}}{16\pi^2}\ln
\left(\frac{e^{-\pi k(T+T^*)}M_*^2}{\mu^2}\right) \, , 
\eea
where $b_A = -3T_A({\rm Adj}) + \sum_\Phi T_A(\Phi)$ ($A=$ UV or IR)
are the one-loop beta function coefficients for the couplings
of gauge fields living only on the branes.

The above expression of low energy brane couplings is what one
would expect based on the notion of position-dependent  
effective cutoff \cite{Randall:2001gb}.
The bare brane couplings $\hat{g}^2_A$ ($A=$ UV or IR) corresponds to
the Wilsonian couplings at the cutoff scale $M_*$ in 
the 5D metric frame of $G_{MN}$.
However in the metric frame of $g_{\mu\nu}$ in which 
$\mu^2=-g^{\mu\nu}\partial_\mu\partial_\nu$ is measured 
(see Eq. (\ref{5dmetric})), $\hat{g}^2_{IR}$ 
can be identified as the Wilsonian coupling at
the effective cutoff scale $e^{-\pi kR}M_*$, while $\hat{g}^2_{UV}$ 
is still the Wilsonian coupling at $M_*$.

As for the low energy bulk gauge couplings, our result
(\ref{4dbulkcoupling}) shows that the (calculable) quantum corrections
are generically of order $\frac{1}{8\pi^2}\ln(M_*^2/\mu^2)
\sim \frac{1}{8\pi^2}\ln(M_{\rm Pl}^2/M_{\rm weak}^2)$.
It reproduces correctly the known 1-loop gauge coupling
in flat  5D orbifold in the limit $kR\rightarrow 0$.
In this flat limit, (\ref{4dbulkcoupling}) is reduced to
\beq
\frac{1}{g_a^2(\mu)}
\,=\,\kappa_a\pi M_*{\rm Re}(T)+
\frac{1}{8\pi^2}\left[\, 
b_a\ln \left(\frac{1}{\mu R}\right)
+b^{\prime\prime}_a\ln\left(M_*R\right)\,\right]+...
%&&\frac{1}{g_{\rm brane}^2}
%\,=\, {\rm Re}(f_{\rm brane})
%+\frac{b_a}{16\pi^2}\ln\left(\frac{M_*^2}{\mu^2}\right)+...
\eeq
where the ellipsis stands for the subleading pieces
of ${\cal O}(1/8\pi^2)$, and
$$
b^{\prime \prime}_a = -2 T_a ({\rm Adj}) +\sum
T_a \left( Q_{UV} \right) 
+ \sum T_a \left( Q_{IR}\right).
$$
This  value of $b_a^{\prime\prime}$ agrees with 
what one would obtain in the explicit KK loop computation \cite{Contino}.
In fact, (\ref{4dbulkcoupling}) can be obtained also
through a totally independent (but more general) 
5D calculation using dimensional regularization \cite{Choi:2002ps}, 
providing a nontrivial check of our result.

The result (\ref{4dbulkcoupling}) is valid for arbitrary value of
$k$ and $\mu$ as long as $k\lesssim M_*$ and $\mu\lesssim M_{KK}$.
When $\pi k (T+T^*)\gg 1$ so that we have the scale hierarchy
(\ref{hierarchy}), one can rewrite (\ref{4dbulkcoupling}) 
in terms of the logarithms of four distinctive mass scales $M_{KK}\approx
\pi ke^{-\pi k(T+T^*)/2}, 1/R$, $k$ and $M_*$.
%as in (\ref{4dcoupling}).
%Since the K\"ahler metric of the massless mode from $H_I$ or $H_I^c$
%severely depends on the kink mass parameter $c_I$,
In this regard, for the hypermultiplets with $\omega_I = \eta_I$
having a massless mode, it is convenient to consider
the following three different classes:
\bea
&&\mbox{Class 1:} \quad \frac{1}{2}-\omega_I c_I \gg \frac{1}{\pi k(T+T^*)}\,,
\nonumber \\
&&\mbox{Class 2:} \quad \left|\frac{1}{2}-\omega_I c_I\right| \ll
\frac{1}{\pi k(T+T^*)}\,,
\nonumber \\
&&\mbox{Class 3:} \quad \frac{1}{2}-\omega_I c_I \ll \frac{-1}{\pi k(T+T^*)}.
\eea
Then (\ref{4dbulkcoupling}) can be written as
\bea
\label{RSrun}
\frac{1}{g_a^2(\mu)} \,&=&\, \kappa_a \pi M_*{\rm Re}(T)+
\frac{b_a}{8\pi^2}\ln\left(\frac{M_{KK}}{\mu}\right)
+\frac{{b}^{\prime}_a}{8\pi^2}\ln\left(\frac{1}{M_{KK}R}\right)
\nonumber \\
&&+\frac{{b}^{\prime\prime}_a}{8\pi^2}\ln\left(kR\right)
+\frac{b^{\prime\prime\prime}_a}{8\pi^2}\ln
\left(\frac{M_*}{k}\right)\, +...,  
\eea
where
\bea
b_a\,&=& \, -3 T_a({\rm Adj}) +\sum T_a(Q_{IR})+\sum T_a(Q_{UV}) +
\sum_{\omega_I = \eta_I} T_a(H_I)\,,
\nonumber \\
{b}^{\prime}_a\,&=&\,-\frac{3}{2}T_a({\rm Adj})+
\sum T_a(Q_{UV})+
\left[\, -\sum_I \eta_I c_I T_a (H_I) +\sum_{\omega_I=\eta_I} T_a(H_I)
\right. \nonumber \\
&& \left. -\sum_{\mbox{class 1}} (1-2 \omega_I c_I) T_a(H_I) \right]\,,
\nonumber \\
{b}^{\prime\prime}_a\,&=& \,-\frac{1}{2}
T_a ({\rm Adj})+\sum T_a(Q_{UV})+
\left[\, -\sum_I \eta_I c_I T_a (H_I) +\sum_{\omega_I=\eta_I} T_a(H_I)
\right. \nonumber \\
&& \left. 
- \sum_{\mbox{class 1}} (1-2\omega_I c_I) T_a(H_I) - \sum_{\mbox{class 2}}
T_a (H_I)\right] \, ,\nonumber \\ 
b^{\prime\prime\prime}_a\,&=&\,-2 T_a({\rm Adj}) + \sum T_a ( Q_{UV})
+\sum T_a (Q_{IR})\,,
\label{RSrunning2}
\eea
and  the ellipsis stands for the subleading pieces
of ${\cal O}(1/8\pi^2)$. 
Obviously  $b_a$ represents the conventional coupling running
at scales between $\mu$ and $M_{KK}$.
On the other hand, other coefficients contain
the KK threshold effects, so their logarithms
can not be interpreted as a coupling running.
%the renormalization effects to make that all of them
%are encoded in logs.

\section{Conclusion}

In this paper, we have examined the
gauge coupling renormalization in orbifold field
theory in the context of 4D effective SUGRA.
The 4D effective SUGRA is a convenient framework to study the gauge
coupling renormalization since the 1-loop gauge couplings
can be determined by the 1-loop induced axion couplings and 
the tree level properties of
4D effective SUGRA which are much easier to be computed.
To be explicit, we take an example of the 5D SUGRA-coupled SYM theory
on a slice of ${\rm AdS}_5$ with four well-seperated mass scales,
the KK threshold scale, the orbifold length, 
the AdS curvature, and the cutoff scale.
In this case, the relevant axion couplings are those of
the graviphoton which can be determined  simply by the chiral anomaly 
structure of the model.
The calculable piece of gauge coupling renormalization
in ${\rm AdS}_5$ is logarithmic, which is generically
of order $\frac{1}{8\pi^2}\ln (M_{\rm Pl}^2/M_{\rm weak}^2)$,
so is  numerically of order unity.
We have calculated such logarithmic corrections in generic 5D SYM models
defined on a slice of ${\rm AdS}_5$ preserving $N=1$ SUSY.

\bigskip
{\bf Note added:}
After this paper is completed, there have appeared several papers
discussing the gauge coupling renormalization in ${\rm AdS}_5$
\cite{Goldberger:2002cz,Agashe:2002bx,Choi:2002zi,Contino:2002kc,Falkowski:new,Choi:2002ps,Goldberger:new,Randall:2002qr}.

\acknowledgments{
HK thanks the particle theory group of University of Washington, Seattle
for their warm hospitality and thanks D. B. Kaplan, A. Katz and A. Nelson
for discussions.
This work is supported in part by the BK21 program of Ministry of Education,
KRF Grant No. 2000-015-DP0080,
the KOSEF Sundo Grant, and the Center for High Energy Physics(CHEP),
Kyungpook National University.
}

\pagebreak
\appendix
\section{5D SUSY transformation and Gauged $U(1)_R$}

In this appendix, we briefly discuss how the $U(1)_R$-gauging
(\ref{u1rcovariant})
is related with the AdS curvature and the mass parameter of the hypermultiplets
through the 5D SUGRA transformation.
Let us first set up the notation for 5D spinor.  
Generic spinor field in 5D SUGRA can be represented by symplectic-Majorana
spinor satisfying
\bea
\bar{\Psi}^i = \Psi_i^\dagger \gamma^0 = (\Psi^i)^T C.
\nonumber
\eea
where
$$
C^T=-C\,,\quad
(\gamma^A)^T=C\gamma^A C^{-1}\,,
$$
and the $SU(2)_R$ doublet index $i=1,2$ is
raised or lowered by $\varepsilon^{ij}=\varepsilon_{ij}=
(i\sigma_2)_{ij}$: 
\bea
\Psi^{i} = \varepsilon^{ij} \Psi _j,\quad 
\Psi_i = \Psi^j \varepsilon_{ji}
\nonumber
\eea
%, \nonumber \\
%\psi^{\alpha i} &=& C^{\alpha \beta} \psi^i_\alpha, \quad
%\psi_\alpha^i = \psi^{\beta i} C_{\beta \alpha}. 
A symplectic-Majorana spinor
contains two
{\em independent} left-handed 4D Weyl spinors $\psi$ and $\psi^c$:
\bea
\Psi^{i=1} = \left(\begin{array}{c} \psi \\ \bar{\psi^c} \end{array}\right),
\quad\quad
\Psi^{i=2} = 
\left(\begin{array}{c} \psi^c \\ -\bar{\psi}\end{array}\right)\,,
\nonumber \eea
so $\Psi^{i=1}$ or $\Psi^{i=2}$
can be regarded as a Dirac spinor.

The 5D SUGRA action of general vector multiplets and hypermultiplets 
and their SUGRA transformations are given in \cite{Ceresole:2000jd}. 
In our notation, the SUGRA transformation of the gravity multiplet
$\{e^A_M,\Psi^i_M,B_M\}$
is given by 
\bea
\label{sugratrans}
\delta_{\xi} e^A_M &=&
\frac{1}{2} \bar{\xi}^i \gamma^A \Psi_{i\,M}\,,\quad
\delta_{\xi} B_M = \frac{i}{2}\overline{\Psi}^i_M\xi_i\, ,
\nonumber \\
%%%%%%%%%%%%%%
\delta_{\xi} \Psi^i_M &=&
D_M \xi^i +
\frac{1}{2} k\epsilon(y) 
(\sigma^3 )^i_{~j} \, e^A_M\gamma_A \xi^j \nonumber \\
&&+ \frac{i}{8} e^A_M
\left( \gamma_{ABC}\xi^i - 4 \eta_{AB} \gamma_C \xi^i\right)
C^{BC} +..., 
%%%%%%%%%%%%%%
\eea
where $C_{MN}=\p_MB_N-\p_NB_M$ and the ellipsis denotes the piece
which is bilinear in hyperino fields.
Comparing $D_M\xi^i$ in $\delta_\xi\Psi^i_M$ with the next term,
one easily finds 
\bea
D_M \xi^i = \nabla_M \xi^i - i \frac{3}{2} k \epsilon(y) 
(\sigma^3)^i_{~j} B_M \xi^i\,, 
\nonumber
\eea
and this determines the $U(1)_R$-gauging for the fields carrying
$SU(2)_R$ index $i$, e.g
\bea
D_M\lambda^{ia}=\nabla_M\lambda^{ia}-i\frac{3}{2}
k\epsilon(y) (\sigma^3)^i_{~j} B_M \lambda^{ja}.
\nonumber
\eea

The SUGRA transformation of hypermultiplet $\left\{ h^i_I, \Psi_I \right\}$
is given by \cite{Ceresole:2000jd}
\bea
\delta_{\xi} h^i_I &=& - i \bar{\xi}^i \Psi_I\,, \nonumber\\
\delta_\xi \Psi_I &=& - \frac{i}{2}\left[\,
\gamma^A \xi^i e^M_A D_M h^j_I\varepsilon_{ji} 
+  k \epsilon(y) \xi^i \left( \frac{3}{2} (\sigma_3)
^j_{~k}-c_I\delta^j_k\right) h^k_I \varepsilon_{ji}\,\right]. 
\label{hyperxform}
\eea
Again comparing the first term in $\delta_\xi\Psi_I$ with the second,
one easily finds
\bea
D_M h^i_I &=& \p_M h^i_I - i k\epsilon(y) \left( \frac{3}{2}(\sigma_3)^i_{~j}
-c_I\delta^i_j \right) k\epsilon(y) B_M h^j_I,
\nonumber
\eea
and so 
\bea
D_M \Psi_I &=& \nabla_M \Psi_I + i c_I k \epsilon(y) B_M \Psi_I.   
\nonumber
\eea

Let us now make sure that the above $U(1)_R$ covariant derivatives
correctly lead to the mass parameters in
$S_{\rm hyper}$ of (\ref{5daction3}).
For the SUGRA transformation (\ref{sugratrans}),
the Killing spinor equation is given by
\bea
D_M \xi^i + \frac{1}{2}k\epsilon(y) (\sigma^3)^i_{~j} e^A_M 
\gamma_A \xi^j = 0, 
\nonumber
\eea
which has the solution
\bea
\xi^{i=1}=\left(\begin{array}{c} \xi \\ 0 \end{array}\right)\,,
\quad\quad
\xi^{i=2}=\left(\begin{array}{c}\ 0\\ \bar{\xi}\end{array}\right)\,,
\nonumber
\eea
where
\bea
\xi (y) = e^{-\frac{1}{2} k |y| (R-3iB_5) }\eta
\label{killingspinor}
\eea
for a constant 4D Weyl spinor  $\eta$ generating
the unbroken $N=1$ global SUSY.

From (\ref{killingspinor}) and (\ref{hyperxform}), one can see that
$\left\{\tilde{h}^1_I, \tilde{\psi}_{I}\right\}$ and 
$\left\{(\tilde{h}^2_I)^*, \tilde{\psi}^c_{I}\right\}$ form 
$N=1$ chiral superfields $H_I$ and $H^c_I$
under the global SUSY generated by
$\eta$, where $\tilde{h}^i_I$, $\tilde{\psi}_I$ and $\tilde{\psi}^c_I$
are defined in (\ref{hyperredef}).
One can also find from (\ref{killingspinor}) and (\ref{hyperxform})
that the $F$-components  of these chiral superfields
are given by
\bea
F_{H_I} &=& \frac{e^{-(1-2c_I)kR|y|}}{R} \partial_5  h^2_I, 
\nonumber\\
F_{H_I^c} &=& -\frac{e^{-(1+2c_I)kR |y|}}{R} \partial_5 \left(h^1_I\right)^*.
\label{Ftermsolution}
\eea
Using the kinetic terms, it is also straightforward to find that 
the D-term action of hypermultiplets is given by 
\cite{Nima,Marti:2001iw}
\bea
\int d^5 x \int d^4 \theta \frac{1}{2}(T+T^*) \left[ 
e^{\left(\frac{1}{2}-c_I\right)k|y| (T+T^*)} H_I H^*_I+
e^{\left(\frac{1}{2}+c_I\right)k|y| (T+T^*)} H^c_I H^{c\,*}_I\right]. 
\nonumber
\eea
In order for the $F$-components of $H_I$ and $H_I^c$  to be given as
(\ref{Ftermsolution}), the $F$-term action should be given by
\bea
\int d^5 x \left[ \int d^2 \theta H^c_I \p_5 H_I + \hbox{h.c.}\right]\,.
\nonumber
\eea
It is then straightforward to see that the above $D$-term and $F$-term actions
give the correct hypermultiplet masses:
\bea
M (\Psi_I) &=& c_I k \epsilon(y),
\nonumber \\
M^2 (h^{i=1}_I) &=& \left(c_I^2 + c_I -\frac{15}{4} \right) k^2 + (3 -2c_I) k 
\left(\delta(y) -\delta(y-\pi)\right), 
\nonumber \\
M^2 (h^{i=2}_I) &=& \left( c_I^2 - c_I- \frac{15}{4} \right) k^2 + (3 +  2c_I) k 
\left(\delta(y) -\delta(y-\pi)\right).  
\nonumber \eea

%%%%%%%%%%%%%%%%%%%%%%%%%%%%%%%%%%%%%%%%%%%%%%%%%%%%%%%%%%%%%%%%%%%%%%%%%%%%%%

\end{document}